\documentclass{iaus}
\usepackage{graphicx}

\title[Star Formation in the LMC] 
{The Conditions for Star Formation at Low Metallicity: Results from the LMC}

\author[Oliveira, van Loon \& Stanimirovic]   
{J.M. Oliveira$^{1}$, J.Th van Loon$^{1}$ \& Stanimirovi\'{c} S.$^{2}$}

\affiliation{$^1$ Astrophysics Group, Keele University, UK; $^2$ Radio Astronomy 
Lab, UC Berkeley, USA}

\pubyear{2006}
\volume{237}  
\pagerange{1--1}
\date{?? and in revised form ??}
\jname{Triggered Star Formation in a Turbulent ISM}
\editors{B. G. Elmegreen \& J. Palous, eds.}

\begin{document}

\renewcommand{\topfraction}{1.}
\renewcommand{\bottomfraction}{1.}
\renewcommand{\textfraction}{0.}

\maketitle

\begin{abstract}
We present our recent work on the conditions under which star formation occurs
in a metal-poor environment, the Large Magellanic Cloud 
([Fe/H]\,$\sim$\,$-$0.4). Water masers are used as beacons of the current star 
formation in H\,{\sc ii} regions. Comparing their location with the dust 
morphology imaged with the Spitzer Space Telescope, and additional H$\alpha$ 
imaging and groundbased near-infrared observations, we conclude that the LMC 
environment seems favourable to sequential star formation triggered by massive 
star feedback (\cite[Oliveira et al. 2006]{oliveira06}). Good examples of this 
are 30\,Doradus and N\,113. There are also H\,{\sc ii} regions, such as N\,105A,
where feedback may not be responsible for the current star formation although 
the nature of one young stellar object (YSO) suggests that feedback may soon 
start making an impact. The chemistry in one YSO hints at a stronger influence 
from irradiation effects in a metal-poor environment where shielding by dust is 
suppressed (\cite[van Loon 2005]{vanloon05}).
\end{abstract}

\begin{figure}[h]
 \centerline{
 \scalebox{0.69}{\includegraphics{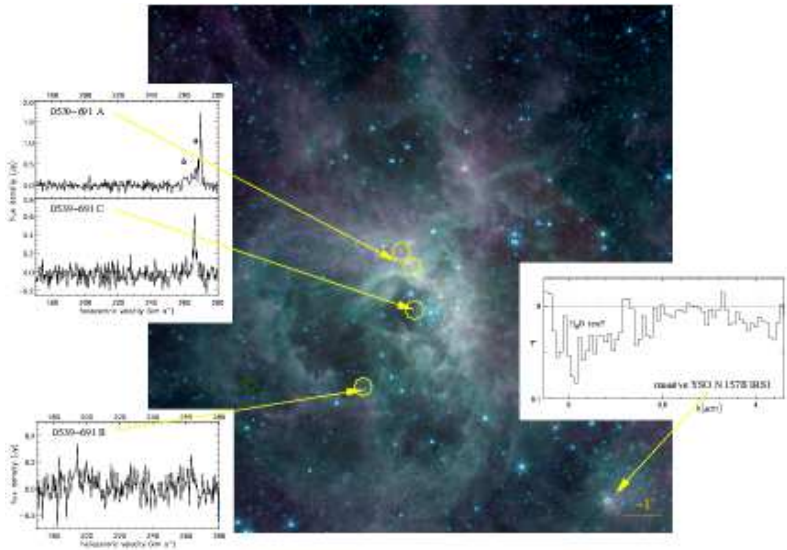}}}
  \caption{Spitzer 3.6\,+\,4.5\,+8\,$\mu$m composite image of 30\,Doradus,
  showing sites of massive star formation. Water masers (l) pinpoint ongoing 
  star formation; the ISAAC/VLT spectrum of the massive YSO (r) shows a hint of
  water ice. More details on this and other H\,{\sc ii} regions in the LMC can be 
  found in \cite{vanloon05} and \cite{oliveira06}.}
\end{figure}


\begin{thebibliography}{}

\bibitem[Oliveira et al. (2006)]{oliveira06}
    {Oliveira J.M. et al.}, 2006, 
     \textit{MNRAS} in press, astro-ph/0609036
\bibitem[van Loon et al. (2005)]{vanloon05}
      {van Loon J.Th. et al.}, 2005,     
     \textit{MNRAS} 364, L71
     
\end{thebibliography}
\end{document}